\documentclass[atoms,communication,accept,moreauthor,pdftex]{mdpi}

\firstpage{1}
\pubvolume{00}
\issuenum{0}
\articlenumber{0}
\pubyear{2019}
\copyrightyear{2019}
\history{Received: 5 March 2019; Accepted: 2 April 2019; Published: 4 April 2019}
\issuenum{0}
\pdfoutput=1

\Title{New Solar Metallicity Measurements}

\Author{Sunny Vagnozzi $^{1,2,\dagger}$}

\AuthorNames{Sunny Vagnozzi}

\address{%
$^{1}$ \quad The Oskar Klein Centre for Cosmoparticle Physics, Department of Physics, Stockholm University, \newline AlbaNova Universitetscentrum, Roslagstullsbacken 21A, SE-106 91 Stockholm, Sweden; sunny.vagnozzi@fysik.su.se\\
$^{2}$ \quad The Nordic Institute for Theoretical Physics (NORDITA), Roslagstullsbacken 23, \newline SE-106 91 Stockholm, Sweden;}

\corres{Correspondence: sunny.vagnozzi@fysik.su.se }

\conference{51st Rencontres de Moriond, \newline Cosmology Session}

\abstract{In the past years, a systematic downward revision of the metallicity of the Sun has led to the ``solar modeling problem'', namely the disagreement between predictions of standard solar models and inferences from helioseismology. Recent solar wind measurements of the metallicity of the Sun, however, provide once more an indication of a high-metallicity Sun. Because of the effects of possible residual fractionation, the derived value of the metallicity $Z_{\odot} = 0.0196 \pm 0.0014$ actually represents a lower limit to the true metallicity of the Sun. However, when compared with helioseismological measurements, solar models computed using these new abundances fail to restore agreement, owing~to the implausibly high abundance of refractory (Mg, Si, S, Fe) elements, which~correlates with a~higher core temperature and hence an overproduction of solar neutrinos. Moreover, the robustness of these measurements is challenged by possible first ionization potential fractionation processes. I~will discuss these solar wind measurements, which leave the ``solar modeling problem'' unsolved.}


\begin{document}

\section{The ``Solar Modeling Problem''}

The metallicity of the Sun, $Z_{\odot}$, i.e., the fraction of solar mass residing in elements heavier than helium, is a fundamental diagnostic of the evolutionary history of our star. Therefore, it is of paramount importance to determine this quantity accurately. Up to 1998, the state-of-the-art was given by the spectroscopic measurements of Anders and Grevesse (AG89) \cite{ag89} and Grevesse and Sauval (GS98) \cite{gs98}, which yielded metallicities of $Z_{\odot} = 0.0202$ and $Z_{\odot} = 0.0170$, respectively. Moreover, heavy element mixtures provided by AG89 and GS98 also yielded good agreement with inferences from helioseismology.

However, following 1998, a systematic downward revision of the solar metallicity has degraded the agreement between models and helioseismology. In particular, the sets of abundances known as AGS05~\cite{ags05} and AGSS09~\cite{agss09} report a metallicity of $Z_{\odot} = 0.0122$ and $Z_{\odot} = 0.0133$, respectively. These~revisions have completely spoiled the previous agreement between models and helioseismology, leading to what is now known as the ``solar modeling problem'' (see, e.g.,~\cite{problem} for a review). The sound speed $u(r)$ is inferred to be $\sim$1\% lower than predicted at the bottom of the convective zone boundary (CZB, a discrepancy of about $10\sigma$), whereas the surface helium abundance $Y_s$ and the CZB $R_b$ are approximately $7\%$ lower and $1.5\%$ higher than those deduced from helioseismology, which amount to discrepancies of approximately $6\sigma$ and $15\sigma$, respectively. Various solutions to the problem have been proposed, including exotic energy transport due to captured dark matter (see, e.g.,~\cite{sarkar,bertone,silk,scott,scott1,scott2}), missing opacity~\cite{krief,mendoza,opacities}, and enhanced convection~\cite{yang}. It is worth pointing out that a revised prediction for the iron opacity at solar interior temperatures hints at a 30--400\% higher opacity than predicted, which goes in the direction of solving the problem, although it only provides half the missing opacity~\cite{bailey}.

In this paper, I will consider another possibility, namely that the solar metallicity determined using spectroscopic methods might in fact not be representative of the true metallicity of the Sun. I will consider alternative measurements from solar wind emerging from polar coronal holes, previously reported in~\cite{vsz16}. The implications that these measurements have for solar models have previously been considered by myself and others in the earlier work~\cite{vfz}. In this communication, I briefly reassess these findings and provide a summary of the current status of solar models in light of solar wind measurements.~\footnote{This communication is based on a talk given at the cosmology session at the 51$^{\text{st}}$ Rencontres de Moriond.}

\section{In Situ Solar Wind Measurements of Metallicity}

All of the heavy element abundance determinations listed above relied on the techniques of photospheric spectroscopy. Despite its broad use within the solar physics community, the interpretation of such measurements is actually far from straightforward. Sophisticated forward modeling taking into account departures from local thermodynamic equilibrium, 3D structures, and radiative transport, as well as accurate knowledge of atomic transition probabilities are required. The extent to which residual systematics affect spectroscopic measurements is not yet fully understood.

There is, nonetheless, a more direct way of measuring the metallicity of the Sun, through {in situ} collection of solar samples. Two current-time sampling techniques exist, which rely on the collection of energetic particles or solar wind samples. We will focus on the latter, i.e., {in situ} solar wind measurements. These types of measurements do not suffer from the difficulties discussed above for spectroscopy. Nonetheless, difficulties and possible systematics exist here as well. For instance, fractionation processes can enhance or deplete the amount of certain ions depending on their ionization and transport histories. By fractionation process, we mean any process that informs about the abundances of nuclei in solar wind samples compared to the photospheric abundances, thus making solar wind samples less representative of the steady state of mass emission from the Sun. Collisional coupling and first ionization potential (FIP) fractionation are among the most important processes at work in this direction. In particular, FIP fractionation refers to the enhancement or depletion of solar wind abundances with respect to photospheric abundance depending on whether the element in question has FIP lower or higher than some reference value, usually taken to be $10\,{\rm eV}$. See, e.g.,~\cite{laming} for a recent comprehensive review on FIP fractionation.

Is it possible to turn this possible weakness into a strength? Fortunately, the answer is yes!
 It has been extensively shown and recently definitely confirmed that fractionation processes are significantly reduced, if not completely absent, in solar wind emerging from polar coronal holes (that is, polar regions near solar minimum) \cite{thomas}. It is possible, and indeed likely, that an important systematics is still at play, namely that we cannot completely exclude a small amount of residual fractionation in these regions. However, it has been shown that any unaccounted residual fractionation would go in the direction of reducing the measured metallicity, and thus, the derived metallicity $Z_{\odot}$ is actually a {lower limit} to the photospheric metallicity~\cite{vsz16,fractionation,fractionation1,
fractionation2,fractionation3,fractionation4}. In view of the recent downward revisions that have led to the ``solar modeling problem'', such a measurement can provide a very important cross-check to the values of metallicity obtained through spectroscopy.

\section{New Measurements: vSZ16}

Very recently, R\"{u}di von Steiger and Thomas Zurbuchen have analyzed data from the ``Solar Wind Ion Composition Spectrometer'' provided by the {Ulysses} mission~\cite{ulysses} to reassess the abundance of heavy elements in the Sun~\cite{vsz16}. In Tab.~\ref{tab1}, we report such values (which we refer to as vSZ16), comparing them to the previous state-of-the-art given by AGSS09. The abundances are reported using the customary logarithmic abundance scale where the abundance of hydrogen is set to be $\epsilon_H = 12.00$. The~fractional variation in abundance for a given element $i$ between vSZ16 and AGSS09 is thus given by $\delta Z_i = 10^{(\epsilon_{_{\text{vSZ16}},i}-\epsilon_{_{\text{AGSS09}},i})}-1$. The total metallicity of the Sun is given by \mbox{$Z_{\odot} = 0.0196 \pm 0.0014$}. In~the following, we will also make a distinction between ``volatile'' and ``refractory'' elements. Usually, volatile elements are those with a low condensation temperature, whereas refractory elements have a~high condensation temperature. In the context of the solar modeling problem in particular, one~usually refers to the elements C, N, O, and Ne as being volatile and to the elements Mg, Si, S, and~Fe as being refractory, adopting a terminology widely used since the seminal work of \mbox{Villante et al.~\cite{villante}}.

\begin{table}[H]
\centering
\caption{Elemental abundances in the vSZ16, AGSS09, and GS98 catalogs and fractional variation between the vSZ16 and AGSS09 catalogs.~Fractional variations between the vSZ16 and GS98 catalogs are not shown since only the AGSS09 catalogue is considered to be the ``baseline'' (and is more widely used by the community currently).}
\label{tab1}%
\begin{tabular}{ccccc}
\toprule
\textbf{Element }&\boldmath {$\epsilon_{_{\text{\textbf{AGSS09}}}}$} & \boldmath$\epsilon_{_{\text{\textbf{GS98}}}}$ &\boldmath $\epsilon_{_{\text{\textbf{vSZ16}}}}$ &\boldmath $\delta Z_i (\text{\textbf{vSZ16}}-\text{\textbf{AGSS09}})$ \\
\midrule
 C & $8.43 \pm 0.05$ & $8.52 \pm 0.06$ & $8.65 \pm 0.08$ & $0.66 \pm 0.15$ \\
 N & $7.83 \pm 0.05$ & $7.92 \pm 0.06$ & $7.97 \pm 0.08$ & $0.38 \pm 0.08$ \\
 O & $8.69 \pm 0.07$ & $8.83 \pm 0.06$ & $8.82 \pm 0.11$ & $0.35 \pm 0.10$ \\
 Ne & $7.93 \pm 0.10$ & $8.08 \pm 0.06$ & $7.79 \pm 0.08$ & $-0.28 \pm 0.08$ \\
 Mg & $7.60 \pm 0.04$ & $7.58 \pm 0.05$ & $7.85 \pm 0.08$ & $0.78 \pm 0.16$ \\
 Si & $7.51 \pm 0.03$ & $7.55 \pm 0.05$ & $7.82 \pm 0.08$ & $1.04 \pm 0.21$ \\
 S & $7.12 \pm 0.03$ & $7.33 \pm 0.11$ & $7.56 \pm 0.08$ & $1.75 \pm 0.35$ \\
 Fe & $7.50 \pm 0.04$ & $7.50 \pm 0.05$ & $7.73 \pm 0.08$ & $0.70 \pm 0.15$ \\
\bottomrule
\end{tabular}
\end{table}

Two comments are in order. First, the variations in the abundance of volatile elements (C, N, O, Ne) is quite contained and actually brings their values close to the previously-accepted values of~\cite{gs98}. The same cannot be said about refractory elements (Mg, Si, S, Fe), for which we see much larger variations, of order $100\%$ or greater. The second comment is required to stress the fact that volatile and refractory elements impact mostly different regions of the solar interior. While volatile elements impact primarily the region around the CZB, refractory elements instead impact mostly the deep interior of the Sun, and in particular, the core. Namely, an increase in the abundance of refractory elements implies a higher core temperature. From this simple point, we can already expect that helioseismological observables, which are mostly sensitive to the conditions around the CZB (such as CZB and sound speed around the CZB), will enjoy an improved agreement, while the disagreement will worsen for observables that are very sensitive to conditions in the core (such as surface helium abundance and neutrino fluxes). In particular, we can expect a huge increase in the neutrino fluxes due to the increase in the abundance of refractories.

\section{Implications for the ``Solar Modeling Problem''}

The changes in elemental abundances listed here directly affect helioseismological observables, which in turn has implications for the ``solar modeling problem''. This occurs because varying the abundance of metals directly affects the radiative opacity of the Sun. Radiative opacity, $\kappa(r)$, describes the coupling between radiation and matter in the hot dense interior of the Sun. In~\cite{vfz}, we worked out the response of four helioseismological observables to the change in metallicity: the sound speed $u(r)$, the surface helium abundance $Y_s$, the convective zone boundary $R_b$, and five solar neutrino fluxes: $\Phi_{\text{pp}}$, $\Phi_{\text{Be}}$, $\Phi_{\text{B}}$, $\Phi_{\text{N}}$, and $\Phi_{\text{O}}$. We did so by making use of the linear solar model (LSM), an alternative to running fully-fledged nonlinear solar codes~\cite{lsm1,lsm2}.

We find that the vSZ16 sound speed represents an improvement over AGSS09 near the CZB boundary, as expected by the change in the abundance of volatile elements, but its disagreement with helioseismology is worsened near the core. Using an appropriately-constructed statistical measure, we argue that the discrepancy between vSZ16 sound speed and helioseismology is at the level of $2.5\sigma$. As for the surface helium abundance, as per expectations due to the increase in the abundance of refractories, we find that this quantity increases well beyond the values allowed by helioseismology, worsening the AGSS09 disagreement. The only helioseismology observable for which vSZ16 abundances represent a large improvement is the CZB, which is compatible with the helioseismology value at $0.88\sigma$, again as expected by inspecting the change in the abundance of volatile elements. Finally, the solar neutrino fluxes increase dramatically in response to the increase in the abundance of refractory elements and subsequent increase of core temperature, leading to fluxes that exceed their measured values by more than $100\%$ (in the case of Be and B neutrino fluxes) or exceed their current upper limits by similar amounts (in the case of N and O neutrinos).

\section{Solar Wind Systematics}

It is clear that the vSZ16 abundances do not solve the ``solar modeling problem'', and this is mostly due to the large increase in the abundance of refractory elements, which leads to unacceptably large values for the surface helium abundance and the neutrino fluxes. Could there be systematics at play that make solar wind measurements an unfaithful representation of the photospheric composition? A comparison between vSZ16 and AGSS09 abundances performed in~\cite{serenelli} suggests that first ionization potential (FIP) fractionation effects are likely at play. The effect of FIP fractionation is to increase the measured abundance of elements whose FIP is greater than that of hydrogen and decrease the measured abundance of elements whose FIP is smaller than that of hydrogen. This would explain the measured high abundance of refractory elements that causes the large disagreement between the model and helioseismology.

Importantly, FIP fractionation can act both in the direction of increasing or decreasing the measured metallicity, unlike the residual sources of fractionation studied by~\cite{vsz16}. This would also invalidate the solar wind measured metallicity being a lower limit to the true metallicity of the Sun. Thus, it is worth going back and re-examining FIP fractionation as a possible systematics to solar wind measurements, and how to reduce it.

\section{Conclusions}

Recent measurements of solar metallicity from solar wind data have provided indication of a high-metallicity Sun, contrary to the systematic downward reassessment in spectroscopic measurements, which has led to the ``solar modeling problem''. The recent determinations by von Steiger and Zurbuchen provide a lower limit (due to possible residual fractionation) on the metallicity of the Sun of $Z_{\odot} = 0.0196$. In this communication, I have discussed how these new measurements improve the agreement with helioseismology only for the sound speed at the bottom of the convective envelope and the convective zone boundary itself, whereas the predictions for the sound speed near the core, the surface helium abundance, and neutrino fluxes are severely discrepant with helioseismological measurements.

The reason is to be searched for within the huge increase in the abundance of refractory elements (Mg, Si, S, Fe), which leads to a hotter core. It could be that the measured values of the abundance of refractories are plagued by systematics due to first ionization potential fractionation, which appears to still be at play. More in-depth studies on the issue are required.

Solar wind provides an exciting complementary probe of solar metallicity. Should first ionization potential fractionation be kept under control, we would be provided with a genuine lower limit to the true metallicity of the Sun. In the meantime, the ``solar modeling problem'' remains unsolved.

\vspace{6pt}

\funding{This research was supported by the Vetenskapsr\r{a}det through Contract No. 638-2013-8993 and the Oskar Klein Centre for Cosmoparticle Physics.}

\acknowledgments{The author wishes to thank the organizers of the 51$^{\text{st}}$ Recontres de Moriond for giving me the opportunity to present this talk, the proceedings of which represent an extended/amended version accounting for advances following the talk. The author greatly benefited from discussions with Katherine Freese, Thomas Zurbuchen, R\"{u}di von Steiger, Mads Frandsen, Subir Sarkar, Ian Shoemaker, Piercarlo Bonifacio, Pat Scott, Aldo Serenelli, and Francesco Villante in preparing this work. Part of this work was conducted at the Michigan Center for Theoretical Physics (now Leinweber Center for Theoretical Physics) at the University of Michigan, Ann Arbor, whom the author thanks for the hospitality.}

\conflictsofinterest{The author declares no conflicts of interest.}

\abbreviations{The following abbreviations are used in this manuscript:\\

\noindent
\begin{tabular}{@{}ll}
AG89 & Anders and Grevesse 1989~\cite{ag89}\\
GS98 & Grevesse and Sauval 1998~\cite{gs98}\\
AGS05 & Asplund, Grevesse and Sauval 2005~\cite{ags05}\\
AGSS09 & Asplund, Grevesse, Sauval and Scott 2009~\cite{agss09}\\
CZB & convective zone boundary\\
vSZ16 & von Steiger and Zurbuchen 2016~\cite{vsz16}\\
FIP & first ionization potential
\end{tabular}}

\reftitle{References}

\end{document}